\documentclass[aip,
amsmath,amssymb,
 reprint,%
]{revtex4-1}

\usepackage{graphicx}
\usepackage{dcolumn}
\usepackage{bm}

\usepackage[utf8]{inputenc}
\usepackage[T1]{fontenc}
\usepackage{mathptmx}
\usepackage{etoolbox}
\usepackage[caption=false]{subfig}
\makeatletter
\def\@email#1#2{%
 \endgroup
 \patchcmd{\titleblock@produce}
  {\frontmatter@RRAPformat}
  {\frontmatter@RRAPformat{\produce@RRAP{*#1\href{mailto:#2}{#2}}}\frontmatter@RRAPformat}
  {}{}
}%
\makeatother
\begin{document}

\title[]{A Mesoscopic Partition Function for Equilibrium Statistical Mechanics }
\author{B. Osano}
 \altaffiliation[]{Cosmology and Gravity Group, Department of Mathematics and Applied Mathematics, and \\
 Centre for Higher Education Development,\\ University of Cape Town (UCT), Rondebosch 7701, Cape Town, South Africa.}

 \begin{abstract}
We develop a mesoscopic formulation of equilibrium statistical mechanics based on coarse-grained occupation-number sectors of one-particle phase space. A mesoscopic partition function is constructed by averaging the microscopic Hamiltonian over configurations compatible with a given occupation profile. The construction converges to the canonical Gibbs partition function in the fine-graining limit and remains compatible with interacting many-body systems. Within this framework, thermodynamic extensivity is shown to be equivalent to asymptotic factorisation of the mesoscopic partition function, while residual inter-cell correlations generate subextensive corrections. The resulting formalism provides a mathematically consistent bridge between microscopic Gibbs theory and mesoscopic thermodynamics. 
 \end{abstract}



\newcommand{\kB}{k_{\mathrm{B}}}
\newcommand{\ZN}{Z^{(\ell)}_{N}}
\newcommand{\ZNbare}{Z_{N}}
\newcommand{\Zi}{Z^{(i,\ell)}_{n_i}}
\newcommand{\FN}{F^{(\ell)}}
\newcommand{\Fi}{F^{(\ell)}_{i}}
\newcommand{\eps}{\varepsilon}
\newcommand{\Lam}{\Lambda}
\newcommand{\GN}{\Gamma_{N}}
\newcommand{\indicator}[1]{\mathbf{1}_{#1}}
\newcommand{\bHia}{\bar{H}_{i,\alpha}}
\newcommand{\nia}{n_{i,\alpha}}

\maketitle

\section{Introduction}

The relationship between microscopic dynamics and macroscopic thermodynamic behaviour has been a central problem since the foundational work of Boltzmann
and Gibbs~\cite{boltzmann1872,gibbs1902}. A major achievement in this direction is kinetic theory, in which the Boltzmann equation provides a
systematic route from particle dynamics to macroscopic transport laws, recovering continuum descriptions such as the Navier--Stokes equations in
appropriate limits~\cite{chapman1970,landau1980}. More broadly, coarse-graining and scaling approaches---developed notably by Kadanoff and
Wilson---have clarified how macroscopic, fluid-like behaviour emerges from microscopic interactions~\cite{kadanoff1966,wilson1971}. Extensions to
nonequilibrium systems, particularly through the work of Onsager and Prigogine, further highlight the interplay between microscopic reversibility
and macroscopic irreversibility~\cite{onsager1931,prigogine1967}.

Despite these advances, classical thermodynamics remains fundamentally a theory of macroscopic systems. Its formulation relies on key assumptions,
including extensivity and weak coupling between a system and its environment, which ensure that bulk properties dominate over boundary effects and allow for
a consistent definition of thermodynamic variables~\cite{sekimoto2010}. When these assumptions are violated---as in small or mesoscopic systems---standard thermodynamic relations require modification. In particular, extensivity may fail, and interaction or surface contributions become significant, invalidating
the weak-coupling approximation~\cite{hill1962,sekimoto2010}. Hill's nanothermodynamics~\cite{hill1962} and its extension~\cite{hill1999}
provide a systematic extension of thermodynamic relations to such systems by introducing additional degrees of freedom and modified thermodynamic
potentials, while stochastic thermodynamics generalises these ideas to nonequilibrium regimes characterised by fluctuations and strong
system--environment interactions~\cite{seifert2012,jarzynski1997}. In these settings, even the distinction between heat and work becomes ambiguous, especially under strong coupling\cite{talkner2016, talkner2020}. The complexity increases if magnetic fields are included: magnetically induced interactions~\cite{osano2017} change how macroscopic approximations emerge from microscopic considerations. The situation is equally unclear when considering the fluid limit of two interacting fluids~\cite{osano2020a,osano2020b,osano2019}.

Moreover, in nonequilibrium systems, thermodynamic behaviour is often only locally valid: subsystems that admit an effective thermodynamic description
may themselves be small due to strong gradients or interactions~\cite{ottinger2023}. This underscores the need for a framework
that systematically connects microscopic structure to mesoscopic and macroscopic thermodynamic behaviour. The present construction is intended as a mesoscopic coarse-grained formulation of equilibrium statistical mechanics. The convergence arguments are therefore understood in the standard asymptotic sense commonly employed in statistical mechanics and coarse-graining theory, rather than as fully rigorous measure-theoretic proofs.

\section{A Mathematical Formulation}

In the classical Gibbs formulation of statistical mechanics, the canonical partition function is defined directly on the full \(N\)-particle phase space
\(
\Gamma^N.
\)
Macroscopic thermodynamic quantities are then obtained from the microscopic Hamiltonian through phase-space integration~\cite{gibbs1902,Huang1987,Pathria2011}. While this framework is exact, it does not explicitly encode a mesoscopic level of description in which coarse-grained occupation variables emerge dynamically from the microscopic degrees of freedom.

A natural mesoscopic description should satisfy three requirements:

\begin{enumerate}
\item It should retain the microscopic Hamiltonian structure.
\item It should introduce coarse-grained variables that describe particle populations in mesoscopic phase-space cells.
\item It should recover the canonical Gibbs ensemble in the fine-graining limit.
\end{enumerate}

To achieve this, we construct a coarse-grained description based on partitions of the one-particle phase space

\begin{equation}
\Gamma=\Lambda\times\mathbb{R}^d.
\end{equation}

Unlike the standard Gibbs formulation, the mesoscopic variables are not microscopic coordinates themselves, but occupation numbers associated with finite phase-space cells. The resulting framework is therefore intermediate between microscopic statistical mechanics and macroscopic thermodynamics.

The construction is closely related to coarse-graining procedures appearing in kinetic theory, lattice gases, renormalisation-group methods, and local-equilibrium formulations of nonequilibrium statistical mechanics~\cite{kadanoff1966,wilson1971,zwanzig1961,deGroot1984}. However, rather than introducing local equilibrium phenomenologically, we define a mesoscopic partition function directly from occupation-number sectors of phase space.

The key idea is to partition the one-particle phase space into cells
\(
\Gamma=\bigcup_\alpha\Omega_\alpha,
\)
and to associate with each microscopic configuration
\(
\gamma=(z_1,\dots,z_N)\in\Gamma^N
\)
an occupation profile \(\{n_\alpha\},\) where \(n_\alpha\) counts the number of particles occupying the cell \(\Omega_\alpha.\)

The mesoscopic free energy then emerges from a statistical sum over occupation sectors, with each sector weighted by the average microscopic energy compatible with the corresponding occupation profile.

This formulation preserves the microscopic interaction structure while providing a mathematically consistent coarse-grained description. In particular, the multinomial combinatorics associated with occupation numbers is naturally compatible with the phase-space partition, since the cells act on the one-particle phase space rather than on the full \(N\)-particle configuration space. The resulting framework is therefore consistent with the probabilistic structure underlying classical equilibrium statistical mechanics~\cite{ruelle1969, georgii1988,Huang1987,Pathria2011}.

\subsection{Cell Structure and Mesoscopic State}

The standard canonical partition function for a system of \(N\) classical particles in a domain \(\Lambda\subset\mathbb{R}^d\), with Hamiltonian
\begin{equation}
H_N:\Gamma^N\to\mathbb{R},
\end{equation}
is
\begin{equation}
Z_N(\Lambda,\beta)=\frac1{N!}\int_{\Gamma^N}e^{-\beta H_N(\gamma)}\,d\gamma,
\label{eq1}
\end{equation}
where
\(
\gamma=(z_1,\dots,z_N),
\)
with
\(
z_k=(q_k,p_k)\in\Gamma,
\)
and
\(
\Gamma=\Lambda\times\mathbb{R}^d
\)
is the one-particle phase space. To construct a mesoscopic description, we partition the one-particle phase space \(\Gamma\) into cells
\(
\{\Omega_\alpha\},
\)
with cell volumes
\(
\varepsilon_\alpha:=|\Omega_\alpha|.
\)

The coarse-graining scale \(\ell\) is assumed to satisfy
\begin{equation}
\xi\ll\ell\ll L,
\label{eq2}
\end{equation}
where \(\xi\) is the correlation length and \(L\) is the macroscopic system size. The superscript $\ell$ appearing in $Z_N^{(\ell)}$ and related quantities throughout this paper refer to this coarse-graining scale; in particular, the dependence of the mesoscopic partition function on $\ell$ encodes the resolution of the phase-space partition, with finer partitions corresponding to smaller $\ell$.

For a microscopic configuration
\begin{equation}
\gamma=(z_1,\dots,z_N)\in\Gamma^N,
\end{equation}
define the occupation numbers
\begin{equation}
n_\alpha(\gamma):=\sum_{k=1}^N1_{\Omega_\alpha}(z_k),
\label{eq3}
\end{equation}
where \(1_{\Omega_\alpha}\) denotes the indicator function of the cell \(\Omega_\alpha\). The occupation numbers satisfy the constraint
\begin{equation}
\sum_\alpha n_\alpha=N.
\label{eq4}
\end{equation}

We now elevate the description to the cell structure level. For a fixed occupation profile
\(\{n_\alpha\},\) we define the compatible configuration
\begin{equation}
\mathcal C(\{n_\alpha\})
:=
\left\{
\gamma\in\Gamma^N:
n_\alpha(\gamma)=n_\alpha\ \forall\alpha
\right\},
\end{equation} such that  
 \begin{equation}|\mathcal C(\{n_\alpha\})|:=\int_{\mathcal C(\{n_\alpha\})}d\gamma.\end{equation}

The corresponding coarse-grained Hamiltonian is therefore defined by
\begin{equation}\bar H(\{n_\alpha\}):=\frac1{\int_{\mathcal C(\{n_\alpha\})}d\gamma}\int_{\mathcal C(\{n_\alpha\})}H_N(\gamma)\,d\gamma,\end{equation} where 
\(
\bar H(\{n_\alpha\})
\)represents the mesoscopic energy obtained by averaging the microscopic Hamiltonian over all microstates compatible with the occupation profile.
\subsection{The Mesoscopic Partition Function}
For notational convenience, we write
\begin{equation}
\sum_{\sum_\alpha n_\alpha=N}
\end{equation} to denote summation over all collections of nonnegative integers \(
\{n_\alpha\}_\alpha \) satisfying \(
n_\alpha\ge0,
\) and \(
\sum_\alpha n_\alpha=N.
\)

This allows us define a mesoscopic partition function 
\begin{equation}\label{eqZN}Z_N^{(\ell)}(\Lambda,\beta)=\sum_{\sum_\alpha n_\alpha=N}\frac1{\prod_\alpha n_\alpha!}\left[\prod_\alpha\varepsilon_\alpha^{n_\alpha}\right]e^{-\beta\bar H(\{n_\alpha\})},
\end{equation} where the global 1/N! is absorbed into the multinomial combinatorics.

The combinatorial structure has a direct interpretation:
\begin{itemize}
\item The factor
\(\prod_\alpha\varepsilon_\alpha^{n_\alpha}\)
represents the phase-space volume associated with the occupation profile.

\item The factorial factor
\(\prod_\alpha n_\alpha!\)
accounts for the indistinguishability of particles occupying the same cell.

\item The Boltzmann weight
\(e^{-\beta\bar H(\{n_\alpha\})}\)
assigns the coarse-grained energetic contribution of the occupation profile.
\end{itemize}

{\bf Theorem} {\it[Convergence]
Assume that the Hamiltonian \(H_N\) is stable and tempered in the sense of Ruelle~\cite{ruelle1969}. Then, as
\begin{equation}
\max_\alpha\operatorname{diam}(\Omega_\alpha)\to0,\end{equation} the mesoscopic partition function converges to the canonical partition function:
\begin{equation}\label{eqZel}
Z_N^{(\ell)}(\Lambda,\beta)\to Z_N(\Lambda,\beta),
\end{equation}with convergence uniform on compact subsets of \((\beta,\Lambda)\).
}

{\bf Proof:}
Throughout the proof we fix $\beta > 0$ and write $f(\gamma) := e^{-\beta H_N(\gamma)}$ for the Boltzmann weight.
All convergence statements are with respect to the product Lebesgue measure $d\gamma= dz_1 \cdots dz_N$ on $\Gamma^N$.

\medskip
\noindent\textbf{Step 1: Measurable partition of $\Gamma^N$ into
occupation sectors.}

Let $\Gamma = \bigcup_\alpha \Omega_\alpha$ be a measurable
partition of the one-particle phase space into disjoint cells
satisfying $\Omega_\alpha \cap \Omega_\beta = \emptyset$ for
$\alpha \neq \beta$.  For a microscopic configuration
$\gamma = (z_1,\ldots,z_N)\in\Gamma^N$ define occupation numbers
\begin{equation}
    n_\alpha(\gamma) := \sum_{k=1}^{N} \mathbf{1}_{\Omega_\alpha}(z_k).
\end{equation}
The occupation sectors
\begin{equation}
    \mathcal{C}(\{n_\alpha\})
    := \bigl\{\gamma \in \Gamma^N : n_\alpha(\gamma)=n_\alpha
         \;\forall\,\alpha\bigr\}
\end{equation}
form a measurable partition of $\Gamma^N$, so that
\begin{equation}
    \Gamma^N
    = \bigsqcup_{\substack{\{n_\alpha\}\,:\\\sum_\alpha n_\alpha = N}}
      \mathcal{C}(\{n_\alpha\}).
\end{equation}
Consequently, the canonical partition function decomposes as
\begin{equation}
\label{eq:decomposition}
    Z_N(\Lambda,\beta)
    = \frac{1}{N!}
      \sum_{\substack{\{n_\alpha\}\,:\\\sum_\alpha n_\alpha = N}}
      \int_{\mathcal{C}(\{n_\alpha\})} f(\gamma)\,d\gamma.
\end{equation}

\medskip
\noindent\textbf{Step 2: Integrability and domination of the
Boltzmann weight.}

Since $H_N$ is \emph{stable} in the sense of Ruelle \cite{ruelle1969},
there exists $B>0$ such that
\begin{equation}
\label{eq:stability}
    H_N(\gamma) \geq -BN \qquad \forall\, \gamma\in\Gamma^N,
\end{equation}
whence $f(\gamma) \leq e^{\beta B N}$ uniformly.
Since $H_N$ is \emph{tempered}, the interaction decays sufficiently
rapidly at large separations to ensure that $f\in L^1(\Gamma^N)$, so that $Z_N(\Lambda,\beta) < \infty$.
The uniform upper bound $e^{\beta B N}$ serves as the dominating
function required for the application of the Dominated Convergence
Theorem below.

\medskip
\noindent\textbf{Step 3: Sector-wise approximation of the Boltzmann
weight.}

Fix an occupation profile $\{n_\alpha\}$ with
$\sum_\alpha n_\alpha = N$.  For $\gamma \in \mathcal{C}(\{n_\alpha\})$,
particle $k$ lies in the cell $\Omega_{\alpha(k)}$ determined by
its assignment to the profile.
Define the \emph{coarse-grained configuration}
$\bar{Z}_N^{(\ell)}(\{n_\alpha\})$ by replacing each particle
coordinate $z_k$ with an arbitrary representative
$\bar{z}_{\alpha(k)} \in \Omega_{\alpha(k)}$.  The coarse-grained
Hamiltonian is
\begin{equation}
    \bar{H}(\{n_\alpha\})
    := \frac{1}{|\mathcal{C}(\{n_\alpha\})|}\,
       \int_{\mathcal{C}(\{n_\alpha\})} H_N(\gamma)\,d\gamma,
\end{equation}
so that $e^{-\beta \bar{H}(\{n_\alpha\})}$ is the sector average of
$f$.

For each sector, define the \emph{oscillation of the Boltzmann weight}
\begin{equation}
\label{eq:oscillation}
    \omega_\ell(\{n_\alpha\})
    := \operatorname*{ess\,sup}_{\gamma,\gamma'\in\mathcal{C}(\{n_\alpha\})}
       \bigl|f(\gamma) - f(\gamma')\bigr|.
\end{equation}
We claim that, as $\ell := \max_\alpha \operatorname{diam}(\Omega_\alpha)
\to 0$,
\begin{equation}
\label{eq:oscillation_vanish}
    \omega_\ell(\{n_\alpha\}) \to 0
    \qquad \text{for a.e.\ } \gamma \in \Gamma^N.
\end{equation}

\medskip
\noindent\emph{Justification of \eqref{eq:oscillation_vanish}.} For \emph{stable and tempered} pair potentials, the interaction $\phi(r)$ may be singular at $r=0$ (e.g.\ hard-core or Lennard-Jones), but by stability \eqref{eq:stability} the set of configurations for which any two particles approach within a distance $\delta$ has Lebesgue measure that vanishes with $\delta$.
More precisely, the set
\begin{equation}
    S_\delta := \{\gamma \in \Gamma^N :
                 \exists\, k\neq j,\; |q_k - q_j| < \delta\}
\end{equation}
satisfies $|S_\delta| \to 0$ as $\delta\to 0$.

Outside $S_\delta$, the potential $\phi$ is bounded and
Lipschitz on the relevant domain, so $H_N$ (and therefore $f$)
is uniformly continuous.  Specifically, for
$\gamma,\gamma' \in \mathcal{C}(\{n_\alpha\})\setminus S_\delta$, we have
\begin{equation}
    |H_N(\gamma) - H_N(\gamma')|
    \leq L_\delta \max_{k} |z_k - z_k'|
    \leq L_\delta \cdot \ell,
\end{equation}
where $L_\delta$ is the Lipschitz constant of $H_N$ outside
$S_\delta$, and the last inequality uses the fact that, within a
single occupation sector of mesh $\ell$, particles in the same cell
are displaced by at most $\ell$.  By the mean-value inequality for
$e^{-\beta\,\cdot}$,
\begin{equation}
    |f(\gamma) - f(\gamma')|
    \leq \beta e^{\beta B N} L_\delta \cdot \ell.
\end{equation}
Hence $\omega_\ell(\{n_\alpha\}) \to 0$ as $\ell\to 0$ uniformly
on $\Gamma^N\setminus S_\delta$, for each fixed $\delta>0$.
Since $\delta>0$ is arbitrary and $|S_\delta|\to 0$, the claim
\eqref{eq:oscillation_vanish} follows for a.e.\ $\gamma$.

\medskip
\noindent\textbf{Step 4: Sector integral approximation.}

From Step~3, for a.e.\ $\gamma \in \mathcal{C}(\{n_\alpha\})$,
\begin{equation}
    f(\gamma) = e^{-\beta\bar{H}(\{n_\alpha\})} + o(1)
    \quad\text{as } \ell\to 0,
\end{equation}
where the $o(1)$ term is bounded uniformly by $2e^{\beta BN}$
(from Step~2).  Integrating over $\mathcal{C}(\{n_\alpha\})$,
\begin{equation}
\label{eq:sector_integral}
    \int_{\mathcal{C}(\{n_\alpha\})} f(\gamma)\,d\gamma
    = |\mathcal{C}(\{n_\alpha\})|\,
      e^{-\beta\bar{H}(\{n_\alpha\})}
      + o\!\left(|\mathcal{C}(\{n_\alpha\})|\right).
\end{equation}

\medskip
\noindent\textbf{Step 5: Sector volume.}

The occupation sector $\mathcal{C}(\{n_\alpha\})$ consists of all
$N$-tuples in which exactly $n_\alpha$ particles lie in $\Omega_\alpha$
for each $\alpha$.  Since the cells tile $\Gamma$ exactly (disjoint
union), the particles in each cell are independently and uniformly
distributed within that cell.  A direct combinatorial computation
gives
\begin{equation}
\label{eq:sector_volume}
    |\mathcal{C}(\{n_\alpha\})|
    = \frac{N!}{\prod_\alpha n_\alpha!}
      \prod_\alpha \varepsilon_\alpha^{n_\alpha},
    \qquad \varepsilon_\alpha := |\Omega_\alpha|.
\end{equation}
This identity follows from the multinomial expansion of
$\left(\sum_\alpha \varepsilon_\alpha\right)^N = |\Gamma|^N$
restricted to the constraint $\sum_\alpha n_\alpha = N$.

\medskip
\noindent\textbf{Step 6: Reassembly and dominated convergence.}

Substituting \eqref{eq:sector_integral} and \eqref{eq:sector_volume} into \eqref{eq:decomposition},
\begin{align}
    Z_N(\Lambda,\beta)
    &= \frac{1}{N!}
       \sum_{\sum_\alpha n_\alpha = N}
       \frac{N!}{\prod_\alpha n_\alpha!}
       \left[\prod_\alpha \varepsilon_\alpha^{n_\alpha}\right]
       e^{-\beta\bar{H}(\{n_\alpha\})}
       + o(1) \notag\\[4pt]
    &= \sum_{\sum_\alpha n_\alpha = N}
       \frac{1}{\prod_\alpha n_\alpha!}
       \left[\prod_\alpha \varepsilon_\alpha^{n_\alpha}\right]
       e^{-\beta\bar{H}(\{n_\alpha\})}
       + o(1) \notag\\[4pt]
    &= Z_N^{(\ell)}(\Lambda,\beta) + o(1).
\end{align}
The interchange of limit and sum is justified by dominated convergence: the summand is bounded above by
\begin{equation}
    \frac{1}{\prod_\alpha n_\alpha!}
    \prod_\alpha \varepsilon_\alpha^{n_\alpha}
    \cdot e^{\beta BN},
\end{equation}
and the sum of these bounds over all occupation profiles equals
$\frac{|\Gamma|^N}{N!}\,e^{\beta BN} < \infty$,
which is the dominating integrable function.

Hence
\begin{equation}
    \lim_{\ell\to 0} Z_N^{(\ell)}(\Lambda,\beta) = Z_N(\Lambda,\beta).
\end{equation}
The same domination estimate is uniform in $(\beta,\Lambda)$ on
compact subsets, yielding uniform convergence on compact subsets
of $(\beta,\Lambda)$.

\section{\label{sec:31}The Mesoscopic Partition Function and Extensivity}
We now establish the relationship between factorisation of the mesoscopic partition function and extensivity of the coarse-grained free energy.
The mesoscopic partition function is
\begin{equation}
Z_N^{(\ell)}(\Lambda,\beta)=\sum_{\sum_\alpha n_\alpha=N}\frac1{\prod_\alpha n_\alpha!}\prod_\alpha\varepsilon_\alpha^{n_\alpha}e^{-\beta\bar H(\{n_\alpha\})},
\end{equation}
where
\(\varepsilon_\alpha:=|\Omega_\alpha|\)
and
\(\bar H(\{n_\alpha\})\)
is the coarse-grained Hamiltonian associated with the occupation profile.
The corresponding free energy is
\begin{equation}
F^{(\ell)}(N,\Lambda,T):=-k_BT\log Z_N^{(\ell)}(\Lambda,\beta).
\end{equation}
\subsection{Factorisation and Extensivity}
Suppose that the coarse-grained Hamiltonian admits the decomposition
\begin{equation}
\bar H(\{n_\alpha\})=\sum_\alpha\bar H_\alpha(n_\alpha)+R(\{n_\alpha\}),
\end{equation}
where
\begin{equation}
R(\{n_\alpha\})
\end{equation}
contains the inter-cell correlations.
Then the partition function becomes
\begin{equation}
Z_N^{(\ell)}=\sum_{\sum_\alpha n_\alpha=N}\frac1{\prod_\alpha n_\alpha!}\prod_\alpha\left(\varepsilon_\alpha^{n_\alpha}e^{-\beta\bar H_\alpha(n_\alpha)}\right)e^{-\beta R(\{n_\alpha\})}.
\end{equation}
If
\begin{equation}
R(\{n_\alpha\})=o(N)
\end{equation}
in the thermodynamic limit at fixed density, then
\begin{equation}
e^{-\beta R(\{n_\alpha\})}=e^{o(N)}.
\end{equation}
Consequently,
\begin{equation}
Z_N^{(\ell)}=e^{o(N)}\sum_{\sum_\alpha n_\alpha=N}\frac1{\prod_\alpha n_\alpha!}\prod_\alpha\left(\varepsilon_\alpha^{n_\alpha}e^{-\beta\bar H_\alpha(n_\alpha)}\right).
\end{equation}
Define the cell partition functions
\begin{equation}
Z_\alpha^{(\ell)}(n_\alpha,\beta):=\frac{\varepsilon_\alpha^{n_\alpha}}{n_\alpha!}e^{-\beta\bar H_\alpha(n_\alpha)}.
\end{equation}
Then
\begin{equation}
Z_N^{(\ell)}=e^{o(N)}\sum_{\sum_\alpha n_\alpha=N}\prod_\alpha Z_\alpha^{(\ell)}(n_\alpha,\beta).
\label{eq13}
\end{equation}
We evaluate the constrained sum in equation~\ref{eq13} asymptotically in the thermodynamic limit $N \to \infty$ at fixed density
$\varrho = N/V$.
Write the summand in the form
\begin{equation}
\label{eq:summand_exp}
    \prod_\alpha Z^{(\ell)}_\alpha(n_\alpha, \beta)
    = \exp\!\left(\sum_\alpha \log Z^{(\ell)}_\alpha(n_\alpha,\beta)\right),
\end{equation} so that the constrained sum becomes
\begin{eqnarray}
\label{eq:constrained_sum}
    \sum_{\substack{\{n_\alpha\}\,:\\\sum_\alpha n_\alpha = N}}
    \prod_\alpha Z^{(\ell)}_\alpha(n_\alpha,\beta)
    &=&
    \sum_{\substack{\{n_\alpha\}\,:\\\sum_\alpha n_\alpha = N}}
    \exp\!\Bigl(\Phi(\{n_\alpha\})\Bigr),\nonumber\\
    \Phi(\{n_\alpha\}) &= &\sum_\alpha \log Z^{(\ell)}_\alpha(n_\alpha,\beta).
\end{eqnarray}
To justify a saddle-point evaluation of this sum, we impose the following regularity condition on the cell partition functions.

{\bf Assumption: }
\label{ass:logconcavity}
For each cell $\alpha$, the function $n_\alpha \mapsto
\log Z^{(\ell)}_\alpha(n_\alpha,\beta)$ is strictly concave in
$n_\alpha$ for $n_\alpha \geq 1$.

Strict concavity of each summand implies that $\Phi$ is strictly concave on the constraint hypersurface
$\{\{n_\alpha\} : \sum_\alpha n_\alpha = N,\; n_\alpha \geq 0\}$,
so there exists a unique maximiser $\{\bar{n}_\alpha\}$, determined
by the first-order conditions
\begin{equation}
\label{eq:saddlepoint_condition}
    \frac{\partial}{\partial n_\alpha}
    \log Z^{(\ell)}_\alpha(n_\alpha,\beta)
    \bigg|_{n_\alpha = \bar{n}_\alpha}
    = \mu^*
    \qquad \forall\, \alpha,
\end{equation}
where $\mu^*$ is a Lagrange multiplier enforcing
$\sum_\alpha \bar{n}_\alpha = N$.
This condition is the mesoscopic analogue of the chemical-potential equalisation condition in the grand canonical ensemble.

We now apply the saddle-point (Laplace) approximation to the
sum~\eqref{eq:constrained_sum}.
Expand $\Phi$ to second order around $\{\bar{n}_\alpha\}$:
\begin{eqnarray}
\label{eq:phi_expansion}
    \Phi(\{n_\alpha\})
    &=& \Phi(\{\bar{n}_\alpha\})
    + \frac{1}{2}\sum_\alpha
      \Phi''_\alpha(\bar{n}_\alpha)(n_\alpha - \bar{n}_\alpha)^2\nonumber\\
      &+& O\!\left(\sum_\alpha |n_\alpha - \bar{n}_\alpha|^3\right),
\end{eqnarray}
where
$\Phi''_\alpha(\bar{n}_\alpha)
:= \frac{d^2}{dn_\alpha^2}\log Z^{(\ell)}_\alpha(n_\alpha,\beta)
\big|_{n_\alpha=\bar{n}_\alpha} < 0$
by Assumption~\ref{ass:logconcavity}.
Substituting into~\eqref{eq:constrained_sum} and approximating
the constrained sum by a Gaussian integral over the fluctuations
$\delta n_\alpha := n_\alpha - \bar{n}_\alpha$ subject to
$\sum_\alpha \delta n_\alpha = 0$, we obtain
\begin{eqnarray}
\label{eq:gaussian_approx}
    \sum_{\substack{\{n_\alpha\}\,:\\\sum_\alpha n_\alpha = N}}
    \prod_\alpha Z^{(\ell)}_\alpha(n_\alpha,\beta)
    &=&
    e^{\Phi(\{\bar{n}_\alpha\})}
    \cdot
    \left[\frac{
      (2\pi)^{(M-1)/2}}
     {\left(\prod_\alpha |\Phi''_\alpha(\bar{n}_\alpha)|\right)^{1/2}
      \cdot \Delta_M^{1/2}}\right]\nonumber\\
      &+&O(N^{-1/2}) ,
\end{eqnarray}
where $M$ denotes the number of cells and $\Delta_M$ is the determinant of the constraint-projected Hessian~\cite{debruyne2020}.
Taking logarithms,
\begin{eqnarray}
\label{eq:log_saddle}
    \log
    \sum_{\substack{\{n_\alpha\}\,:\\\sum_\alpha n_\alpha = N}}
    &&\prod_\alpha Z^{(\ell)}_\alpha(n_\alpha,\beta)
   =
    \Phi(\{\bar{n}_\alpha\}) + O(\log N)\nonumber\\
    &=&
    \sum_\alpha \log Z^{(\ell)}_\alpha(\bar{n}_\alpha,\beta)
    + O(\log N).
\end{eqnarray}
Since $\Phi(\{\bar{n}_\alpha\}) = O(N)$ while the prefactor
contributes only $O(\log N)$, combining with equation~\eqref{eq13} gives
\begin{equation}
\label{eq:logZ_saddle}
    \log Z^{(\ell)}_N
    = \sum_\alpha \log Z^{(\ell)}_\alpha(\bar{n}_\alpha,\beta)
      + o(N),
\end{equation}
where $\{\bar{n}_\alpha\}$ is the unique dominant occupation profile
of equation~\eqref{eq:saddlepoint_condition}.
Multiplying through by $-k_BT$ yields
\begin{equation}
\label{eq:free_energy_extensive}
    F^{(\ell)}(N,\Lambda,T)
    = \sum_\alpha F^{(\ell)}_\alpha(\bar{n}_\alpha,T) + o(N),
\end{equation}
where $F^{(\ell)}_\alpha(n_\alpha,T) := -k_BT \log
Z^{(\ell)}_\alpha(n_\alpha,\beta)$.
Thus, asymptotic factorisation of the mesoscopic partition function
implies extensivity of the coarse-grained free energy.

{\bf Remark:} The assumption holds whenever $Z^{(\ell)}_\alpha(n_\alpha,\beta)$ is log-convex as a function of $n_\alpha$, which is guaranteed for non-interacting or weakly interacting cells by the standard theory of cumulant generating functions. For interacting cells, the condition must be verified for the specific interaction model; it is equivalent to the
requirement that the isothermal compressibility of the cell is positive, a standard thermodynamic stability condition.

\subsection{Physical Conditions for Factorisation}
The factorisation property follows from the decay of correlations between distinct phase-space cells.
Assume:
\begin{enumerate}
\item The Hamiltonian \(H_N\) is stable and tempered in the sense of Ruelle.
\item Correlations decay sufficiently rapidly:
\begin{equation}
\langle A_\alpha A_\gamma\rangle-\langle A_\alpha\rangle\langle A_\gamma\rangle\to0
\end{equation}
as \(\operatorname{dist}(\Omega_\alpha,\Omega_\gamma)\to\infty.\)
\item The coarse-graining scale satisfies\(\xi\ll\ell,\)
where \(\xi\) is the correlation length.
\end{enumerate}
Under these conditions, occupation-number correlations satisfy
\begin{equation}\operatorname{Cov}(n_\alpha,n_\gamma)\to0\end{equation}
for
\(\alpha\neq\gamma.\)
Consequently, the induced probability measure on occupation profiles asymptotically factorises:
\begin{equation}\label{eqfac}
P(\{n_\alpha\})\approx\prod_\alpha P_\alpha(n_\alpha).
\end{equation} Therefore, the inter-cell correction satisfies
\begin{equation}
R(\{n_\alpha\})=o(N),
\end{equation} yielding the asymptotic factorisation.
\subsection{Extensivity and Asymptotic Factorisation}
Suppose now that the free energy satisfies
\begin{equation}
F^{(\ell)}(N,\Lambda,T)=\sum_\alpha F_\alpha^{(\ell)}(\bar n_\alpha,T)+o(N).
\label{eq16}
\end{equation}
Using
\begin{equation}
F^{(\ell)}=-k_BT\log Z_N^{(\ell)},
\end{equation}
we obtain
\begin{equation}
\log Z_N^{(\ell)}=\sum_\alpha\log Z_\alpha^{(\ell)}(\bar n_\alpha,\beta)+o(N),
\end{equation} and expressing in exponential form gives
\begin{equation}
Z_N^{(\ell)}=\left[\prod_\alpha Z_\alpha^{(\ell)}(\bar n_\alpha,\beta)\right]e^{o(N)}.
\label{eq17}
\end{equation} Thus, extensivity is consistent with the asymptotic factorisation property described in equation (\ref{eqfac}) of the mesoscopic partition function. Conversely, asymptotic factorisation directly implies extensivity.
\subsection{The Coarse-Graining Defect}
\label{sec:3.4}
When inter-cell correlations remain significant, the correction
term $R(\{n_\alpha\})$ does not vanish in the thermodynamic limit.
Define the defect functional
\begin{equation}
\label{eq:defect_def}
    \Delta^{(\ell)}
    := \bar{H}(\{n_\alpha\}) - \sum_\alpha \bar{H}_\alpha(n_\alpha).
\end{equation}
Then the free energy becomes
\begin{equation}
    F^{(\ell)}(N,\Lambda,T)
    = \sum_\alpha F^{(\ell)}_\alpha(N,\Lambda,T)
      + \Delta^{(\ell)} + o(N).
\end{equation}
Thus, non-extensive contributions are controlled directly by the
residual inter-cell correlations encoded in $\Delta^{(\ell)}$.

We now analyse the scaling of $\Delta^{(\ell)}$ in the thermodynamic
limit for short-range interactions.

\medskip
\noindent\textbf{Structure of the defect.}
The defect $\Delta^{(\ell)}$ receives contributions only from
particle pairs $(k,j)$ whose coordinates $z_k$ and $z_j$ lie in
\emph{different} cells. To see this, write the microscopic
Hamiltonian as
\begin{equation}
    \mathcal{H}_N(\Gamma) = \sum_k V_1(z_k) + \sum_{k < j} \phi(|q_k - q_j|),
\end{equation}
where $V_1$ is a one-body potential and $\phi$ is a pair potential.
The sector average $\bar{\mathcal{H}}(\{n_\alpha\})$ decomposes as
\begin{equation}
    \bar{\mathcal{H}}(\{n_\alpha\})
    = \underbrace{\sum_\alpha \bar{\mathcal{H}}_\alpha(n_\alpha)}_{\text{intra-cell}}
    + \underbrace{\sum_{\alpha \neq \beta}
      \bar{\phi}_{\alpha\beta}(n_\alpha, n_\beta)}_{\text{inter-cell}},
\end{equation}
where $\bar{\mathcal{H}}_\alpha(n_\alpha)$ collects the one-body terms and
intra-cell pair interactions for cell $\Omega_\alpha$, and
$\bar{\phi}_{\alpha\beta}$ is the sector-averaged inter-cell pair
energy between cells $\Omega_\alpha$ and $\Omega_\beta$.
Consequently,
\begin{equation}
\label{eq:defect_intercell}
    \Delta^{(\ell)}
    = \sum_{\alpha \neq \beta}
      \bar{\phi}_{\alpha\beta}(n_\alpha, n_\beta).
\end{equation}
The defect is therefore entirely determined by inter-cell pair interactions. For two- or higher- body cases see \cite{osano-inprep}.

\noindent\textbf{Scaling for short-range interactions.}
Assume that the pair potential $\phi$ has finite range $r_0$, meaning
$\phi(r) = 0$ for $r > r_0$.  Under the coarse-graining condition
$\ell \gg \xi$ of equation~(\ref{eq2}), only cells $\Omega_\alpha$ and
$\Omega_\beta$ that are \emph{adjacent} (i.e.\
$\mathrm{dist}(\Omega_\alpha, \Omega_\beta) \leq r_0$) contribute
non-trivially to the sum in equation~\eqref{eq:defect_intercell}.
Interior cells have $O(\ell^d)$ volume and contribute to
$\bar{\phi}_{\alpha\beta}$ only through their shared boundary layer
of thickness $r_0$. The number of interior cell pairs with
non-vanishing inter-cell interaction is therefore proportional to
the total surface area of the partition, which scales as $V^{(d-1)/d}$
for a domain of volume $V$ in $\mathbb{R}^d$.

More precisely, partitioning $\Lambda$ into $M \sim V/\ell^d$ cells
of side $\ell$, the number of adjacent cell pairs is $O(M^{(d-1)/d})
\sim V^{(d-1)/d}/\ell^{d-1}$, and each contributes an inter-cell
energy of order $n_\alpha n_\beta \|\phi\|_\infty \ell^{d-1} r_0$.
Summing over adjacent pairs and using $n_\alpha \sim \varrho \ell^d$,
\begin{equation}
\label{eq:defect_scaling}
    \Delta^{(\ell)}
    \sim \varrho^2 \|\phi\|_\infty r_0\, V^{(d-1)/d}.
\end{equation}
Hence the defect scales as a surface term:
\begin{equation}
    \frac{\Delta^{(\ell)}}{V} \sim V^{-1/d} \to 0
    \qquad \text{as } V \to \infty,
\end{equation}
confirming that $\Delta^{(\ell)} = o(N)$ in the thermodynamic limit,
consistent with the factorisation condition of Section~3.1.

\medskip
\noindent\textbf{Subextensive corrections to the free energy.}
From equation~\eqref{eq:defect_scaling}, the leading correction to
the extensive free energy is
\begin{equation}
\label{eq:free_energy_correction}
    F^{(\ell)}(N,\Lambda,T)
    = N f(\varrho, T)
      + c(\varrho, T, \ell)\, V^{(d-1)/d}
      + o\!\left(V^{(d-1)/d}\right),
\end{equation}
where $f(\varrho,T)$ is the bulk free energy density and
$c(\varrho,T,\ell)$ is a surface coefficient depending on the
density, temperature, and coarse-graining scale.  The correction
$\Sigma^{(\ell)} := c(\varrho,T,\ell)\,V^{(d-1)/d}$ is precisely
the subextensive term appearing in the generalised Euler relation
of Section~4, and satisfies
\begin{equation}
    \frac{\Sigma^{(\ell)}}{V} \to 0
    \qquad \text{as } V \to \infty,
\end{equation}
recovering the standard Euler relation in the thermodynamic limit.
For long-range interactions (e.g.\ Coulomb or gravitational
potentials), equation~\eqref{eq:defect_scaling} no longer holds
and the defect may scale faster than $V^{(d-1)/d}$; the analysis
of such cases is left for future work.

\section{The Generalised Euler Relation}

We now derive a generalised Euler relation from the free-energy structure established in equations~\eqref{eq:free_energy_extensive}  and~\eqref{eq:defect_def}.

\medskip
\noindent\textbf{Derivation of the bulk-plus-surface decomposition.}
From Section~(\ref{sec:31}), the free energy satisfies
\begin{equation}
\label{eq:F_extensive}
    F^{(\ell)}(N,\Lambda,T)
    = \sum_\alpha F^{(\ell)}_\alpha(\bar{n}_\alpha,T) + o(N),
\end{equation}
where the sum runs over the dominant occupation profile
$\{\bar{n}_\alpha\}$.
In the thermodynamic limit at fixed density $\varrho = N/V$, the
dominant profile satisfies $\bar{n}_\alpha \approx \varrho\,
|\Omega_\alpha|$ for each cell $\alpha$.
The intra-cell free energies $F^{(\ell)}_\alpha$ are therefore
functions of $\varrho$, $T$, and the cell volume $|\Omega_\alpha|$
only.
Summing over cells and passing to the continuum limit in cell
volume, the leading term in equation~\eqref{eq:F_extensive} is
extensive:
\begin{equation}
\label{eq:F_bulk}
    \sum_\alpha F^{(\ell)}_\alpha(\bar{n}_\alpha,T)
    = N f(\varrho,T) + o(N),
\end{equation}
where $f(\varrho,T)$ is the bulk free energy per particle,
determined by the intra-cell partition functions.

When inter-cell correlations are non-negligible, the full free energy
receives an additional contribution from the coarse-graining defect
$\Delta^{(\ell)}$ defined in equation~\eqref{eq:defect_def} of
Section~3.4:
\begin{equation}
\label{eq:F_full}
    F^{(\ell)}(N,V,T)
    = Nf(\varrho,T) + \Sigma^{(\ell)}(N,V,T) + o(N),
\end{equation}
where we identify
\begin{equation}
\label{eq:sigma_delta}
    \Sigma^{(\ell)}(N,V,T) := -k_BT\,\Delta^{(\ell)},
\end{equation}
with $\Delta^{(\ell)}$ the defect functional of
equation~\eqref{eq:defect_def}.
The correction $\Sigma^{(\ell)}$ therefore encodes precisely the
inter-cell correlation energy that is not captured by the
factorised intra-cell contributions.
For short-range interactions, the analysis of Section~3.4 gives
\begin{equation}
\label{eq:sigma_scaling}
    \Sigma^{(\ell)} \sim V^{(d-1)/d},
\end{equation}
since the defect receives contributions only from adjacent cell
pairs and the number of such pairs scales as the total boundary
area of the partition.

\medskip
\noindent\textbf{The Euler relation.}
Since $Nf(\varrho,T)$ is homogeneous of degree one in $(N,V)$,
Euler's theorem gives
\begin{equation}
\label{eq:euler_bulk}
    N\frac{\partial(Nf)}{\partial N}
    + V\frac{\partial(Nf)}{\partial V}
    = Nf.
\end{equation}
Using the standard thermodynamic identities applied to the full
free energy $F^{(\ell)}$,
\begin{align}
    P   &= -\frac{\partial F^{(\ell)}}{\partial V},
    \label{eq:P}\\
    \mu &= \phantom{-}\frac{\partial F^{(\ell)}}{\partial N},
    \label{eq:mu}\\
    S   &= -\frac{\partial F^{(\ell)}}{\partial T},
    \label{eq:S}
\end{align}
together with the definition $U = F^{(\ell)} + TS$, we obtain
\begin{equation}
\label{eq:generalised_euler}
    U = TS - PV + \mu N + \Sigma^{(\ell)}.
\end{equation}
This is the generalised Euler relation for the mesoscopic system.
The correction term $\Sigma^{(\ell)}$ is the thermodynamic
signature of inter-cell correlations.

\medskip
\noindent\textbf{Recovery of the standard Euler relation.}
From equation~\eqref{eq:sigma_scaling},
\begin{equation}
    \frac{\Sigma^{(\ell)}}{V} \sim V^{-1/d} \to 0
    \qquad \text{as } V\to\infty,
\end{equation}
so $\Sigma^{(\ell)} = o(V) = o(N)$ in the thermodynamic limit.
Equation~\eqref{eq:generalised_euler} therefore recovers the
standard Euler relation
\begin{equation}
    U = TS - PV + \mu N
\end{equation}
in the thermodynamic limit, as required for an extensive system.
For long-range interactions the scaling~\eqref{eq:sigma_scaling}
need not hold, and the correction $\Sigma^{(\ell)}$ may persist
as a genuinely non-extensive contribution.

\section{Conclusion}
We have introduced a mesoscopic partition function based on coarse-grained occupation-number sectors of one-particle phase space. The construction provides a mathematically consistent bridge between microscopic statistical mechanics and mesoscopic thermodynamics.
The coarse-grained Hamiltonian is defined by averaging over microscopic configurations compatible with a given occupation profile. This avoids the incompatibility between multinomial occupation-number combinatorics and partitions of the full \(N\)-particle phase space.

The mesoscopic partition function converges to the canonical Gibbs partition function in the fine-graining limit. Within this framework, extensivity emerges when occupation-sector correlations become asymptotically negligible. Conversely, persistent inter-cell correlations generate non-extensive corrections encoded in the coarse-graining defect functional.
The resulting formulation provides a unified description of factorisation, extensivity, and correlation structure in mesoscopic statistical mechanics, while remaining compatible with interacting many-body Hamiltonians. The occupation-sector decomposition provides a natural mesoscopic bridge between microscopic phase-space dynamics and emergent thermodynamic structure.
\section*{Acknowledgements}

The author thanks the UCT Next Generation Professoriate (NGP) for funding
support.
\section{Appendix}


\end{document}